\documentclass[twocolumn]{emulateapj}

\makeatletter
\usepackage{etoolbox}
\patchcmd\H@refstepcounter{\protected@edef}{\protected@xdef}{}{}
\makeatother

\shorttitle{Jupiter's formation location}
\shortauthors{\"Oberg \& Wordsworth}

\begin{document}

\title{Jupiter's composition suggests its core assembled exterior to the N$_2$ snowline}

\email{koberg@cfa.harvard.edu}

\author{Karin I \"Oberg}
\affil{Harvard-Smithsonian Center for Astrophysics, 60 Garden St, Cambridge, MA 02138, USA}

\author{Robin Wordsworth}
\affiliation{Harvard Paulson School of Engineering and Applied Sciences, Harvard University, Cambridge, MA 02140, USA}
\affiliation{Department of Earth and Planetary Sciences, Harvard University, Cambridge, MA 02140, USA}

\begin{abstract}

Jupiter's atmosphere is enriched in C, N, S, P, Ar, Kr and Xe with respect to solar abundances by a factor of $\sim$3. Gas Giant envelopes are mainly enriched through the dissolution of solids in the atmosphere, and  this constant enrichment factor is puzzling since several of the above elements are not expected to have been in the solid phase in Jupiter's feeding zone; most seriously, Ar and the main carrier of N, N$_2$, only condense at the very low temperatures, 21--26~K, associated with the outer solar nebula. We propose that a plausible solution to the enigma of Jupiter's uniform enrichment pattern is that Jupiter's core formed exterior to the N$_2$ and Ar snowlines, beyond 30~au, resulting in a Solar composition core in all volatiles heavier than Ne. During envelope accretion and planetesimal bombardment, some of the core mixed in with the envelope causing the observed enrichment pattern. We show that this scenario naturally produces the observed atmosphere composition, even with substantial pollution from N-poor pebble and planetesimal accretion in Jupiter's final feeding zone. We note that giant core formation at large nebular radii is consistent with recent models of gas giant core formation through pebble accretion, which requires the core to form exterior to Jupiter's current location to counter rapid inward migration during the core and envelope  formation process. If this scenario is common, gas giant core formation may account for many of the gaps observed in protoplanetary disks between 10s and 100 au. 

\end{abstract}


\section{Introduction} \label{sec:intro}

The compositions of planets are linked to the chemical conditions in the solar nebula. Since chemical conditions change across the Nebula, a planet's composition provides clues to its formation locations, and therefore to its dynamical past \citep{Dodson-Robinson09,Oberg11d,Ciesla15}. Of all known giant planets, Jupiter presents the most well-constrained composition because of {\it in situ} measurements by the Galileo and Juno missions \citep{Niemann96,Bolton17}. Importantly for this paper, the Galileo mission revealed that C, N, S, P, Ar, Kr and Xe are all enriched  with respect to hydrogen compared to Solar abundances \citep{Owen99}.  Whether O is enriched is unknown -- Galileo recorded a sub-Solar O abundance, but this probably does not reflect the bulk O abundance. 
Gas giant envelopes can become enriched through a number of processes --  core erosion and mixing, accretion of enriched gas, and dissolution of accreting pebbles and planetesimals during envelope accretion or in the subsequent clean-up stage \citep{Hueso03,Estrada17} -- and some enrichment compared to Solar abundances is therefore not surprising. What is surprising, however, is that all the above species are enriched by approximately the same factor, $\sim$3.

This enrichment pattern is surprising because the expected solar nebula solid composition at $\sim$3--5~au, the assumed formation location of Jupiter in most models \citep[e.g.][]{Gomes05}, is decidedly non-Solar, and most models explain Jupiter's enrichment by solid accretion and dissolution . At 3--5~au, the solids are expected to have been mainly composed of refractory material and water ice \citep{Ciesla15}, and therefore rich in oxygen (O), sulfur (S) and phosphor (P) \citep{Anders89,Asplund09}, but comparatively poor in carbon (C), and very poor in nitrogen (N) and noble gases (Ar, Kr and Xe), because important carbon carriers (CO$_2$ and CO), and the dominant nitrogen carrier (N$_2$), as well as Xe, Kr, and Ar only freeze out further out in the solar nebula. Accretion of such solids by a gas giant would enrich its envelope strongly in O, S and P, slightly in C, and not at all in N, Ar, Kr, and Xe, in tension with the observed uniform factor of 3 enrichment in all observed elements. 

A possible solution to the presented tension is entrapment of hyper-volatiles in water ice, which could maintain more C, and some N, Ar, Kr and Xe in solids at 5~au. Indeed several models have invoked clathration or entrapment of hypervolatiles in amorphous water ice as explanations of Jupiter's enrichment pattern \citep{Lunine85,Owen99,Gautier01,Hersant04, Gautier05, Mousis09}. These models face some difficulties, however. First, they require large amounts of water ice to entrap all other volatiles. This should result in an excess enrichment in oxygen by a factor of a few compared to other elements in Jupiter's atmosphere, for which there is so far no evidence.  Second, entrapment of CO, N$_2$ and Ar through clathration requires low nebular temperatures \citep{Lunine85,Gautier01}, 40~K and less, which is difficult to achieve in the Nebula at 3--5~au if radiative heating is taken into account. 

A possible solution to the difficulty of locally entrapping hyper-volatiles at 3--5~au, is radial drift of cold pebbles from the outer solar nebula into Jupiter's feeding zone \citep[e.g.][]{Cuzzi04,Oberg16b}. A similar idea underpins a recent study by \citet{Mousis19}, who considers sublimation of entrapped volatiles in inward-drifting amorphous water ice pebbles. The enriched gas is then accreted onto Jupiter. While this process likely plays a role it should not lead to a uniform enhancement in e.g. N and C.
First, even at low temperatures, N$_2$ entrapment is inefficient compared to CO entrapment \citep[e.g.][]{Bar-Nun85,Bar-Nun07,Yokochi12}, which may explain low N$_2$ abundances in comets \citep{Cochran00,Cochran02,Rubin15}. Second, the nebular model must be fine-tuned to result in a pebble population at 3--5~au that originates exclusively from the cold, outer solar nebula region, rather than from a range of radii, most of which would not allow for efficient N$_2$ entrapment.

A simpler explanation to Jupiter's enrichment pattern is that Jupiter's core formed in the outer solar system, beyond the N$_2$ and Ar snowlines, from solids with Solar ratios of all elements heavier than Ne. During envelope accretion, and planetesimal and embryo impacts some of the core was then mixed in with the envelope causing the observed enrichment pattern. Such a formation scenario may appear implausible at first sight, but is supported by both recent theory and observations. First, recent models of core formation through pebble accretion in actively accreting disks only produce a Jupiter-sized planet at Jupiter's location if the core forms substantially further out, at nebular radii $>$15~au \citep{Bitsch15,Bitsch19,Pirani19}. In earlier generations of gas giant formation models, gas giant formation was limited by long core formation and envelope accretion timescales \citep[e.g.][]{pollack1996formation,Hueso03}, which typically exceeded 5~Myrs at 5~au. This is longer than the typical 2-3 Myr lifetime of observed protoplanetary disks \citep{Mamajek09}, and since timescales increase with nebular radius, gas giant formation in the outer Solar nebula seemed excluded. By contrast planet core formation through pebble accretion is fast, and in recent models the whole gas giant formation process -- core formation, envelope accretion, and inward migration -- can be completed in $<$1Myr. Even if core formation begins at $\sim$40~au the complete process takes only $\sim$2 Myrs \citep{Bitsch15} 

Second, Millimeter observations of analogs to our solar nebula, i.e. of protoplanetary disks, have revealed that gaps appear common at disk radii of 10--100~au \citep{Andrews18,Huang18}. These gaps are proposed to be associated with actively forming planets, and while other explanations exist, there is at least one example where the there is supporting kinematic evidence for protoplanets in the disk gaps \citep{Teague18,Pinte18}. If these gaps are indeed carved out by planets, the gap widths and depths can be used to constrain planet masses and this was recently done for the DSHARP disk sample \citep{Andrews18,Zhang18}. The result is that the gaps can be explained by planets and planetary embryos of masses between $\sim$ 10 Earth masses and a few Jupiter masses, suggestive of that gas giant, or at least gas giant cores often begin their existence at large disk radii. 

The pebble accretion scenario has been used in one study to explore whether it can indeed explain Jupiter's composition \citep{Ali-Dib17}, using a full-scale planet formation and migration model. They found, that core formation in the outer solar nebula could not alone account for Jupiter's nitrogen enrichment, probably because of the location of the N$_2$ snowline in their nebular model. In this paper we take a simpler, toy model approach to explore expected enrichment patterns in Jupiter's envelope when its core forms in the outer regions of the solar nebula, with outer regions defined with respect to the N$_2$ and Ar snowline locations. \S \ref{sec:model} introduces the nebular snowline model used throughout the paper. In \S \ref{sec:results} we present our fiducial enrichment model and explore how sensitive it is to core and envelope formation locations, as well as to pollution during the clean-up stage. We discuss the results in \S \ref{sec:disc} before offering some brief conclusions in \S \ref{sec:conc}.

\section{Solar Nebula Model} \label{sec:model}

\subsection{Density and Temperature Structure}

To explore the link between Jupiter's core formation location and its observed envelope composition, we construct a simple, static toy model of the radial composition of solids and gas in the solar nebula midplane.
 We follow the common assumption of radial power laws in surface density and temperature \citep[e.g.][]{Lewis74,Chiang10}:

\begin{eqnarray}
    \Sigma_{\rm H} = \Sigma_{\rm H,1au}\left(\frac{r}{1{\rm \: au}}\right)^{-\gamma}\\
    n_{\rm H} = \frac{\Sigma_{\rm H} }{\sqrt{2 \pi}H},\\
    H = \frac{k_{\rm b}T}{m_{\rm H}\mu}\sqrt{\frac{r^3}{G M_{\rm Sun}}},\\
    T_{\rm mid} = T_{\rm mid,2au}\left(\frac{r}{2 {\rm \: au}}\right)^{-q},
\end{eqnarray}
where $\Sigma_{\rm H}$ is the column density, $r$ is the disk radius in au, $\Sigma_{\rm H,1au}$ is the column density  at 1~au, which we set to 1500 g cm$^{-2}$, and $\gamma$ is the surface dense powerlaw index, which is typically assumed to be $3/2$ for the solar nebula \citep{Chiang10}. The hydrogen nuclei density $n_{\rm H}$ is calculated from the surface density, and the isothermal scale height $H$. Finally, $T_{\rm mid}$ is the midplane temperature, and $q$ is the temperature powerlaw index. The latter is expected to be $\sim3/7$ in the outer disk (exterior to a few au), where reprocessed Solar radiation dominates disk heating \citep{Chiang10}. However, higher values of up to 0.7 have been inferred from observations \citep{Andrews07}. 

Rather than adopting theoretical estimates of $T_{\rm mid}$ and $g$, we use data on solar nebula H$_2$O, CO and N$_2$ snowline locations to set their values. The water snowline has been localized to $\sim2$~au, though the location likely evolved with time as Jupiter was forming \citep{Min10}. The CO and N$_2$ snowlines are more uncertain. Based on comet compositions the CO snowline was likely located in the comet forming zone, since comets present a large diversity of CO abundances \citep{Mumma11}. This fits with recent estimates of the CO snowline in the TW Hya disk \citep{Zhang17}. Appreciable amounts of N$_2$ in comets are rare \citep{Cochran00,Cochran02,Rubin15}, though a N$_2$/CO ratio of 0.15 was recently reported in one comet \citep{Cochran18}, and the majority of comets therefore likely formed interior to the N$_2$ snowline, placing the N$_2$ snowline in the outer range of the proposed comet-forming region of 5 and 35~au \citep{Mumma11}. 
Finally, Pluto appears to be rich in N$_2$, which would be consistent with formation exterior to the N$_2$ snowline, though other explanations have been given as well \citep{Stern18}. Pluto likely formed at 20--30~au \citep{Kenyon12}, and we therefore tune our disk model temperature profile such that the N$_2$ snowline is at 20--30~au. 

\begin{figure}
    \centering
    \includegraphics[scale=0.6]{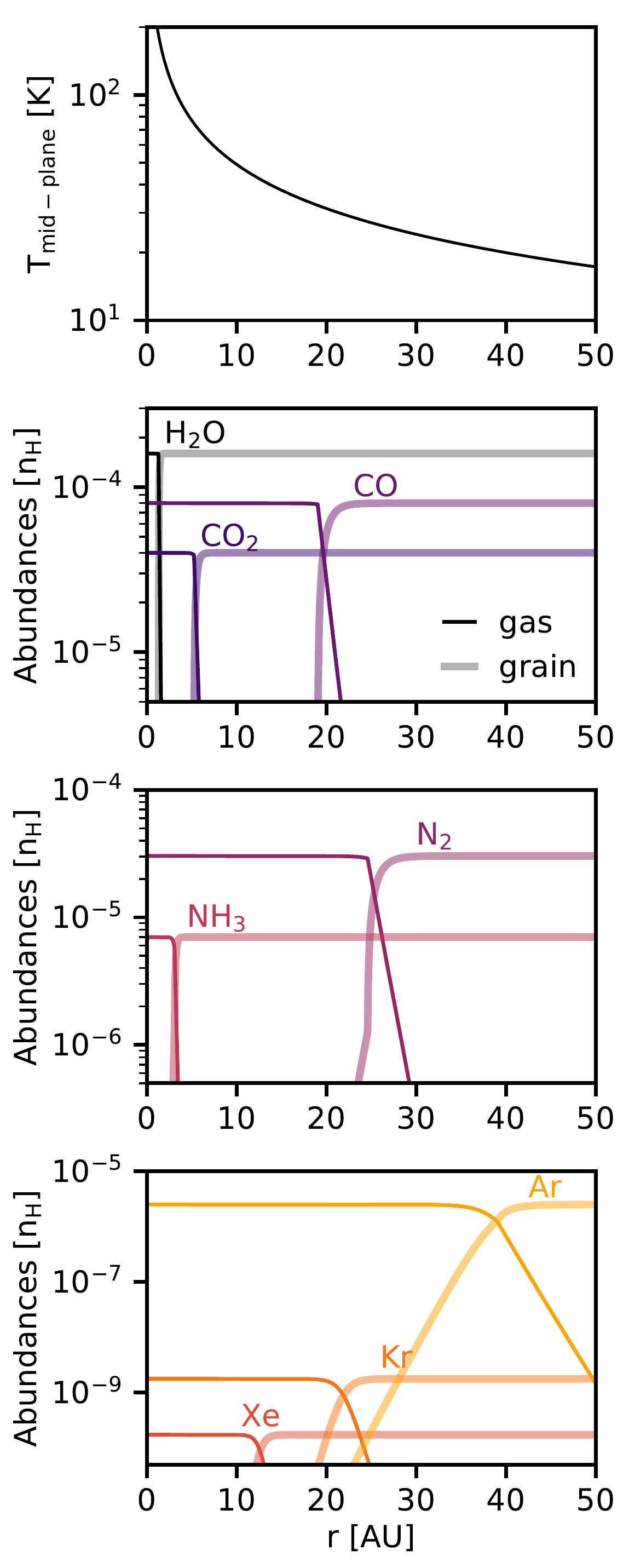}
    \caption{{\sl Top:} Adopted temperature profile for the solar nebula midplane. {\sl Lower three panels:}  Abundances and snowline locations of major carriers of O, C, and N, respectively, as well as of the noble gases assuming no entrapment in less volatile ices.}
    \label{fig:snowlines}
\end{figure}

Using standard values for H$_2$O, CO and N$_2$ sublimation energies of 5800, 1180, and 1051~K, respectively, where the latter two values assume CO and N$_2$ sublimation from a water-rich ice, we obtain a reasonable fit to the above snowline constraints when $T_{\rm mid,2au}=140$~K and $q=0.65$. The resulting temperature profile  is shown in the top panel of Fig \ref{fig:snowlines}, and the snowline locations of H$_2$O, CO and N$_2$ are plotted in the panels below. 
We note, however, that a warmer or more shallow temperature profile would have been inferred if pebble drift was included in the model, since pebble drift move snowlines inward compared to the static case \citep[e.g.][]{Piso15}. The presented temperature profile should therefore be viewed as a convenient tool to estimate gas and solid abundances across the solar nebula rather than a an accurate model of the solar nebula thermal structure.

\subsection{Molecular Abundances and Snowline Locations}

The gas and solid-state distributions of elements in a disk is primarily set by the condensation lines of major element carriers, e.g. silicate grains, H$_2$O CO and CO$_2$ for oxygen. In this paper we consider only the elements that have been quantified in Jupiter's atmosphere, i.e. O, C, N, P, S, Xe, Kr and Ar. The carriers of several of these are unfortunately poorly constrained and the estimates in Table \ref{tab:carriers} should be treated as provisional. Our general strategy is to use data from the interstellar medium (ISM), which provides abundance baselines for he young solar nebula, and augment with solar system data as available.

In the dense ISM, oxygen is mainly carried by silicate grains, H$_2$O, CO and CO$_2$, with similar amounts of O in the first three carriers, while CO$_2$ is present at a $\sim25\%$ level compared to water \citep{Whittet10,Oberg11c,Boogert15}.  In comets, H$_2$O appears more abundant than silicate, while CO$_2$ is $\sim$20\% compared to water and CO abundances vary \citep{Mumma11}. We adopt a mixed scheme with equal abundances of O in H$_2$O and silicate grains, and CO and CO$_2$ abundances that are 50\% and 25\% compared to H$_2$O respectively, and set the absolute abundances so that they add up to Solar \citep{Asplund09}. 

Based on the oxygen budget above, 50\% of the carbon is in CO and CO$_2$. We split the remaining carbon into one part volatile organics, using ethane as a model system, and three parts carbon grains and refractory organics. For nitrogen, ISM data shows that $\sim$10\% of nitrogen is in NH$_3$, while solar system, ISM and protoplanetary disk data alike indicate that most of the remaining ($\sim$90\%) nitrogen is in N$_2$ \citep{Pontoppidan14,Oberg11c}, and we use these estimates. Both S and P are heavily depleted in the ISM, indicative of refractory carriers. Finally the noble gases are assumed to be present in atomic form. All abundances are listed in Table \ref{tab:carriers}.

We assume silicate grains, carbon grains and refractory organics, S and P, which are all present in the solid state at all relevant radii. The abundances of all other species are modeled to solely depend on the changing balance between sublimation and condensation as a function of radius. Condensation rates are described by:

\begin{equation}
    R_{\rm cond,i}=n_{\rm gr}n_{i,gas}v_{\rm i}\sigma_{\rm gr},
\end{equation}
where $n_{\rm gr}$ is the grain number density, $n_{i,gas}$ the number density of species $i$ in the gas-phase, $v_{\rm i}$ the collisional velocity of grains and species $i$ which is assumed to the be the thermal velocity of $i$, and $\sigma_{\rm gr}$ is the collisional cross section which is the cross section of the grain. It is useful to rewrite this equation in terms of abundances $x$ with respect to the main constituent of the solar nebula, hydrogen nuclei, in which case the equation instead becomes:

\begin{equation}
    R_{\rm cond,i}=x_{\rm gr}x_{i,gas}n_{\rm H}^2v_{\rm i}\sigma_{\rm gr}.
\end{equation}

Sublimation is described in detail by e.g. \citet{Fraser01} and \citet{Bisschop06}. In summary, for low surface coverages where all surface molecules are available for sublimation, sublimation is calculated from:

\begin{eqnarray}
    R_{\rm subl,i}= x_{\rm i,grain}n_{\rm H}\times\nu_{\rm i}{\rm Exp}(E_{\rm subl,i}/T) ,
\end{eqnarray}
where $x_{\rm i,grain}$ is the abundance of $i$ frozen out on grains, $\nu_{\rm i}$ is the attempt frequency, $E_{\rm subl,i}$ is the sublimation barrier in units of Kelvin, and $T$ is the grain temperature which is assumed to be perfectly coupled to the gas temperature in the dense disk midplane. Sublimation energies and attempt frequencies for all species are listed in Table \ref{tab:carriers}.  For higher surface coverages, where only the top layer of the ice can sublime, the sublimation rate is instead:

\begin{eqnarray}
    R_{\rm subl,i}= x_{\rm grain}n_{\rm H}10^{15}4\sigma_{\rm grain}\times\nu_{\rm i}{\rm Exp}(E_{\rm subl,i}/T). 
\end{eqnarray}

We assume that there is a steady state between sublimation and condensation at each radius, and that the total abundance $x_{\rm i} = x_{\rm i,gas}+x_{i,gr}$ is constant. We then use the adopted disk temperature and midplane hydrogen density profile to calculate the gas and grain abundance of each species as a function of solar nebula radius. The results of these calculation are shown in Fig. \ref{fig:snowlines}. Note that the CO snowline is at $\sim$20~Au, and the N$_2$ snowline at $\sim$26~au in agreement with the above snowline location constraints from solar system composition data. The Ar snowline at $\sim$40~au is the most distant one in our model; under nebular conditions and using the binding to water ice reported by \citet{Smith16}, Ar only freezes out $<$21~K. More recent, unpublished data suggests a slightly higher Ar condensation temperature of $\sim$25~K (Schneiderman, private communication), which would move the Ar snowline close to the N$_2$ snowline. We use the published value in this study, but note that the existing Ar sublimation data may place Jupiter's inception 10~au further out than is actually required.

\begin{table}[]
    \centering
    \begin{tabular}{l c c c}
    \hline
    \hline
        Molecule & Abundance $x_{\rm i}$ [$n_{\rm H}$] & $\nu_{\rm i}$ [s$^{-1}$] & $E_{\rm subl,i}$ [K]   \\
        \hline
        H$_2$O&1.6$\times10^{-4}$&$4\times10^{13}$& 5800$^1$\\
        CO$_2$&4$\times10^{-5}$&$1\times10^{13}$& 2700$^2$\\
        CO&8$\times10^{-5}$&$7\times10^{11}$& 1180$^3$\\
        Volatile organics&3$\times10^{-5}$&$6\times10^{16}$& 2500$^4$\\
        N$_2$&3$\times10^{-5}$&$8\times10^{11}$& 1050$^3$\\
        NH$_3$&7$\times10^{-6}$&$1\times10^{13}$& 3800$^5$\\
        Ar&2.5$\times10^{-6}$&$6\times10^{11}$& 870$^{6,7}$\\
        Kr&1.8$\times10^{-9}$&$1.2\times10^{14}$& 1380$^6$\\
        Xe&1.7$\times10^{-10}$&$4.6\times10^{14}$& 1970$^6$\\
        \hline
    \end{tabular}
    \caption{Adopted molecular abundances, desorption attempt frequencies and energies}
    \label{tab:carriers}
    $^1$ \citet{Fraser01}, $^2$ \citet{Sandford90}, $^3$ \citet{Fayolle16}, $^4$ \citet{Behmard19}, $^5$ \citet{Suhasaria15}, $^6$ \citet{Smith16}, using their binding energies to compact water.
\end{table}

\section{Results} \label{sec:results}

\subsection{Nebular Elemental Solid Ratios}

\begin{figure}
    \centering
    \includegraphics[scale=0.6]{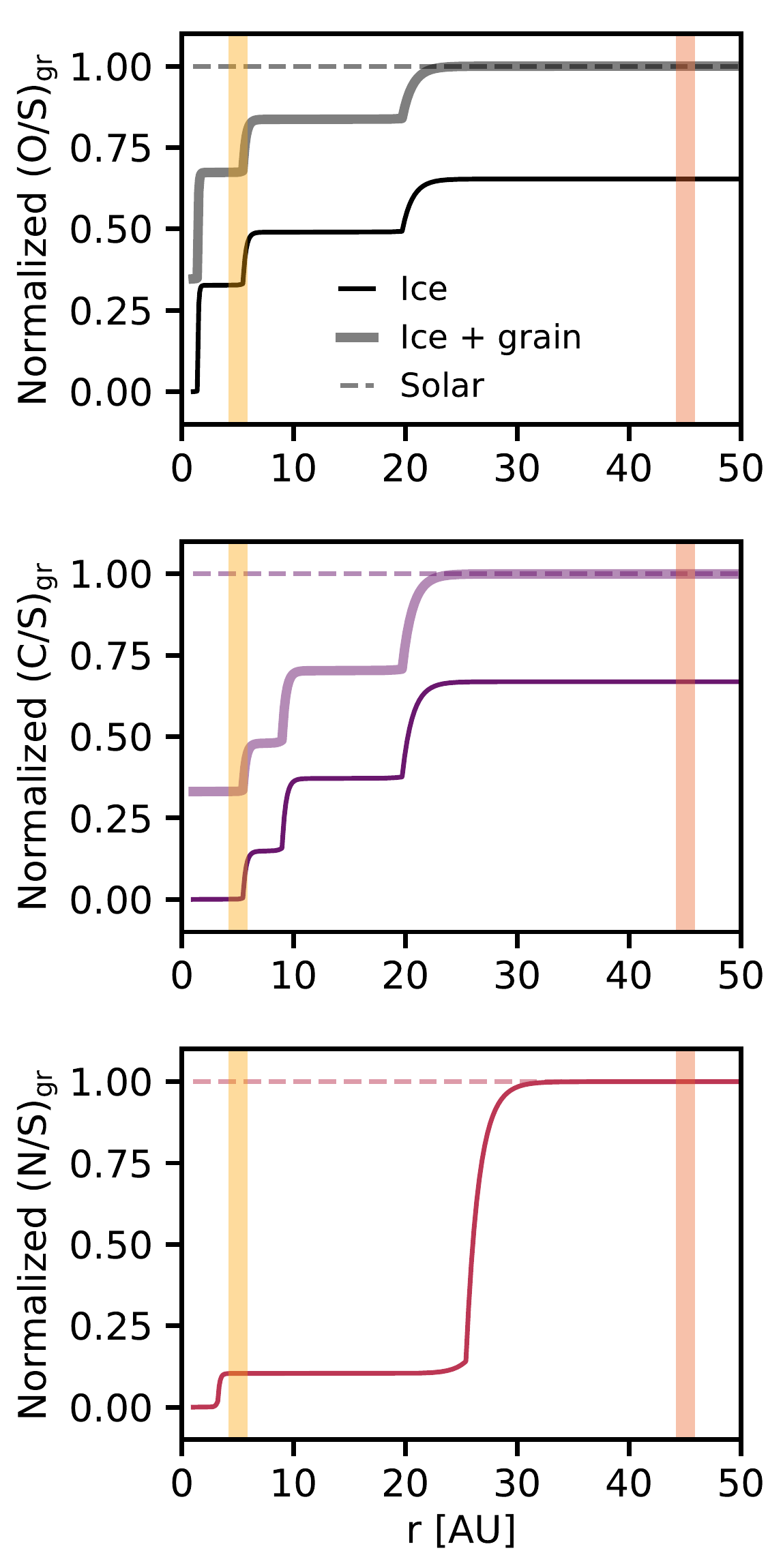}
    \caption{Expected abundance ratios between O, C, N and S in the solar nebula normalized to Solar. Thin lines account for ices, and thick lines for ices and refractory grain material. The transparent light orange band marks the current location of Jupiter at 5~au, and the transparent dark orange band the location of its core formation at 45~au in our fiducial model, though core formation at 30~au would be sufficient to account to Jupiter's N enrichment. Note that without substantial entrapment in water ice the C/S and N/S ratios in solids are low around Jupiter's current location.}
    \label{fig:elements}
\end{figure}

We can use the snowline calculations above to calculate the relative abundances of O, C, N, P, S, Ar, Kr and Xe in solids, and hence the achievable enrichment patterns for a planet forming at different radii. Figure \ref{fig:elements} shows the O, C and N solid abundances with respect to S, normalized to Solar abundances. Sulfur is a reasonable reference element since it is a solid at all relevant radii. At 5~au, the solid-state O/S, C/S and N/S ratios are 0.76, 0.41, and 0.10 respectively. This entails that an atmosphere enriched with solids at this radii, i.e. the current location of Jupiter, will not obtain a close to uniform enrichment pattern unless entrapment of N$_2$ and noble gases is extremely efficient. O/S, C/S and N/S solid-state ratios approach Solar as the distance from the Sun increases, and become Solar beyond the N$_2$ snowline at 26~au. Beyond 40 au, even Ar freezes out resulting in icy solids with Solar composition except for in in H, He and Ne.

\subsection{Fiducial Model for Jupiter's Formation}

\begin{figure}
    \centering
    \includegraphics[scale=0.7]{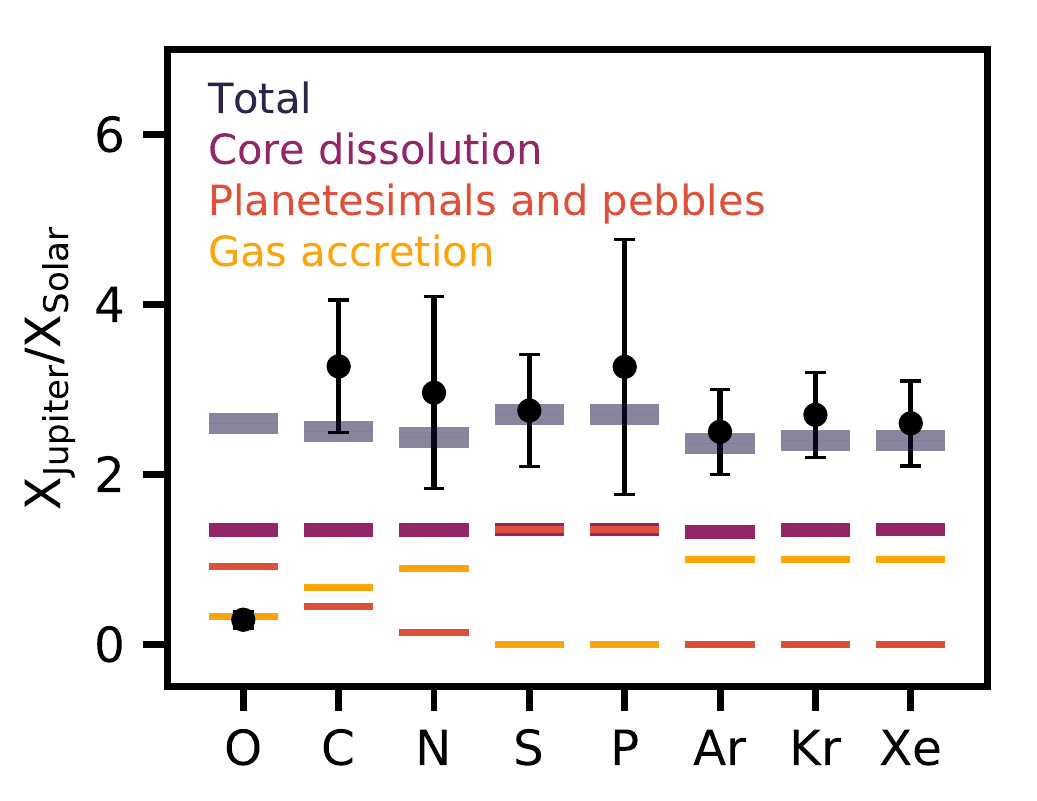}
    \caption{Expected elemental enhancement ratios in Jupiter's envelope in our fiducial model compared to Solar (thick blue lines), and the contributions from core mixing, planetesimal dissolution, and gas envelope accretion. Black points are measurements from Galileo (O, C, S, Ar, Kr and Xe \citep{Mahaffy00,Wong04}), Cassini (P \citep{Fletcher09}), and Juno (N \citep{Bolton17}).}
    \label{fig:jupiter-fid}
\end{figure}

Considering 1) that solids are only expected to have Solar composition in O, C, N, S, P and noble gases in the outer solar nebula, and 2) that Jupiter appears uniformly enriched in these same elements, it follows that Jupiter likely formed with substantial amounts of outer Nebula solids. To account for this, our fiducial model assumes that Jupiter's core formed beyond 45~au, and thus contains a large reservoir of heavy elements at Solar ratios. 
Based on e.g. \citet{Bitsch15}, we further assume that the newly formed core migrated inwards toward its current location, while it accreted most of its envelope -- the estimated formation+migration timescale from 45~au is $\sim$2.5~Myrs \citep{Bitsch15}. For simplicity we assume that the majority of the envelope was accreted close to 5~au and thus has a 5~au gas composition. Over time this envelope became enriched by dissolution and outward mixing of the core (dredging) \citep{stevenson1982formation,wahl2017comparing}, and by local (5~au) accretion of solids which dissolved in the gaseous envelope.

The model hinges on the possibility of dissolution of both the core and later accreted solids. Models of impacts of ice-rich planetesimals with radii of 30 m to 1 km show complete ablation in the outer envelope for a wide range of parameters
\cite{pollack1996formation,iaroslavitz2007atmospheric}. More refractory planetesimals and larger objects show more mixed behavior \citep{Pinhas16,liu2019formation}. To our knowledge there is no similar calculation for pebbles, but it is commonly assumed that icy pebbles completely dissolve, while refractory pebbles may or may not reach the core \citep{Venturini16}. For simplicity, we assume complete dissolution of impacting pebbles and planetesimals in the gaseous envelope, but this may require an update as more calculations become available. The main effect on our model if refractory material does not dissolve would be to lower the P and S enhancements originating from impacting pebbles and planetesimals. The efficiency of core mixing is even more uncertain and we discuss it further in \S\ref{sec:disc}.

The relative contributions of core mixing and dissolution of locally accreted solids to Jupiter's atmosphere are unknown. In our fiducial model we scale their relative contributions, such that half of the sulfur in Jupiter's envelope, our reference species, originates from core mixing, and half from solid accretion at 5~au. This 50-50 divide is somewhat arbitrary and simply encodes a scenario where there is substantial contributions from both reservoirs (in the next section we explore scenarios where one or the other dominates). We scale the total core and local enrichment such that the sulfur enrichment in Jupiter's envelope agrees with observations. The resulting elemental composition in Jupiter's atmosphere can then be traced back to three different sources: mixing of the core, which results in a constant enhancement of all species, accretion of gas at 5 au, which is sub-solar in O, S and P, and almost solar in all other elements, and accretion of solids at 5~au, which are rich in O, S and P, contain some C, little N, and no noble gases.

Figure \ref{fig:jupiter-fid} shows the resulting enrichment pattern. The model agrees with all measured abundances, except for oxygen, but the Galileo measurement of oxygen is generally assumed to not be representative of Jupiter's true composition. For all elements, core mixing provides at least 50\% of the measured envelope abundances given our model assumptions, while local solid and gas accretion provides the remainder; gas accretion is more important for C, N and noble gases, while local solid accretion accounts for the remainder of O, S and P.

\subsection{Model grid results}

\begin{figure*}
    \centering
    \includegraphics[scale=0.65]{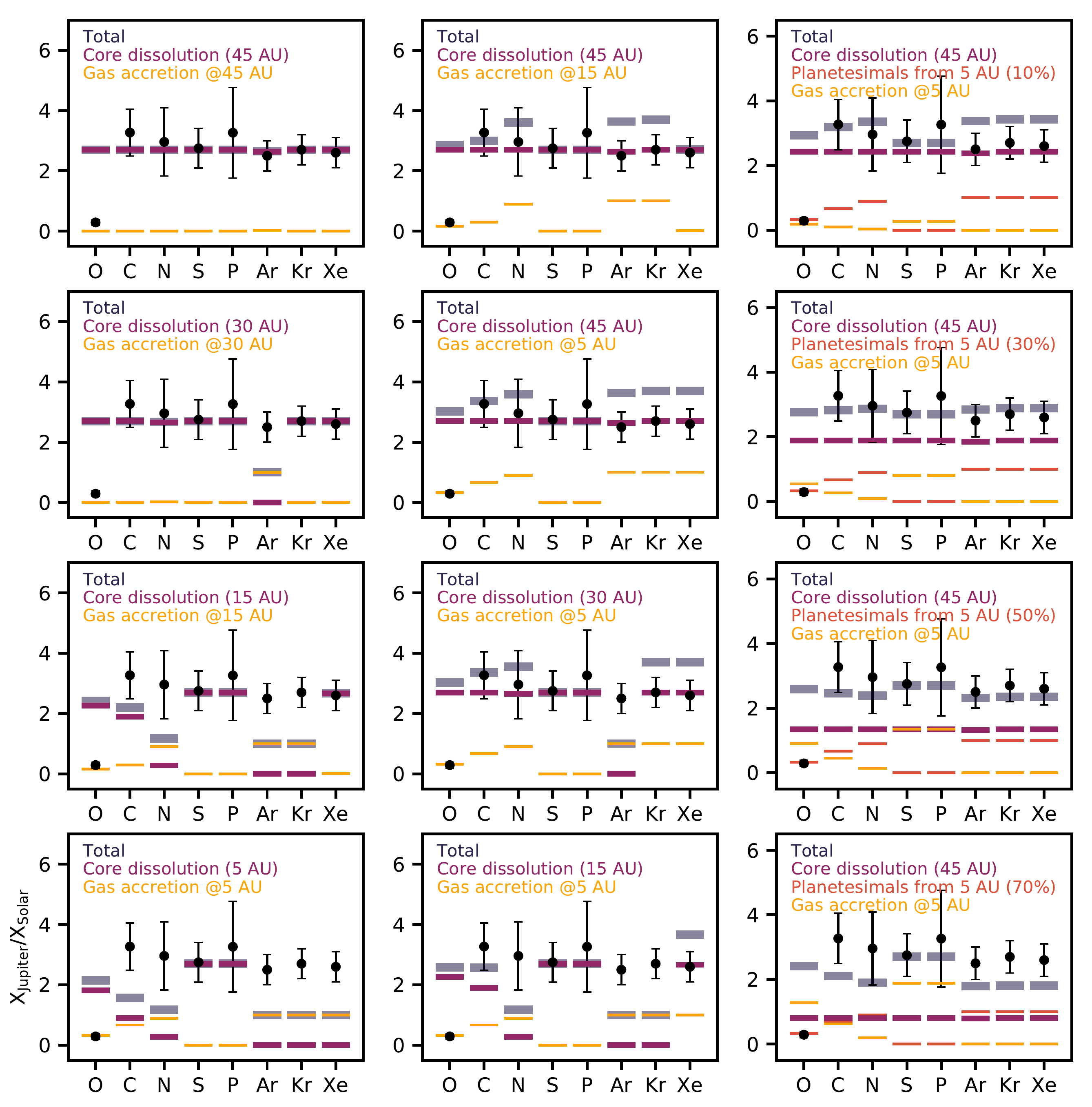}
    \caption{Same as Fig. \ref{fig:jupiter-fid}, but for different combinations of solid and gas accretion locations. {\it Left:} Predicted elemental enhancements in Jupiter's envelope when solids and envelope are accreted at the same disk radius. {\it Middle:} Predicted elemental enhancements when the gas is accreted interior to the core formation location. {\it Right:} Predicted enhancements when a fraction of the solids dissolved in the enveloped originates from a core formed at 45~au and a fraction from planetesimals accreted at 5~au. Note that only scenarios where Jupiter's core forms at 45~au fit all data. If Ar condenses out closer to the N$_2$ snowline, as suggested by recent unpublished data, core formation beyond 30~au is sufficient to explain observations.}
    \label{fig:jupiter-grid}
\end{figure*}

To evaluate the sensitivity of our results to the formation and migration history of Jupiter, Figure \ref{fig:jupiter-grid} explores the outcomes of different combinations of core formation locations, gas envelope accretion locations, and relative contributions from core mixing and planetesimal dissolution on the predicted element enhancements in Jupiter's envelope.  Similarly to the fiducial model, we fix the total amount of solid dissolution such that the sulphur enhancement matches the observed one.

In the first set of models (left column), the core and gas are accreted at the same location between 45 and 5~au, and all enhancements can be traced back to the solid and gas compositions at the initial formation location. This is a toy version of a scenario where initial migration is slow. In this scenario, formation at 45~au is consistent with data because all elements in question are at equal abundances exterior to the Ar snowline. Complete formation at 30~au also provides a good fit to the data with the exception of Ar, which as mentioned above may in reality condense further in than this model suggests. Formation at smaller radii provides a poor fit, because of the low solid abundance of N, Ar and Kr at 15~au, and of N, C, Ar, Kr and Xe at 5~au, which entails that they are predicted to be present in Jupiter's atmosphere at Solar levels, rather than the $\sim$2 times higher levels observed. While complete planet formation beyond 30~au is consistent with Jupiter's abundance pattern, we note that it is not supported by either theory \citep[e.g.][]{Bitsch15} or by observations \citep[e.g.][]{Kruijer17} and are only presented here as model end members.

In the second column of Fig. \ref{fig:jupiter-grid} we consider models where the solids originate in the outer solar system, while the gas envelope is accreted at 15 or 5~au. This mimics scenarios where the core forms early at 15-45~au, and the envelope is accreted during inward migration closer to Jupiter's present-day location, and later planetesimal accretion and dissolution is inefficient. None of the models result in perfect fits to observations. The models where the core forms at 30-45~au provide good fits to C, N, S and P, but not to the noble gases, which are either under- or overpredicted, while core formation at 15~au also fails to reproduce N.

The third and last column of Fig. \ref{fig:jupiter-grid} explores elemental enhancements in Jupiter's envelope when a fraction of the dissolved solids polluting the envelope originates from a core formed at 45~au, and the remainder from solid accretion at 5~au. This assumes the same scenario as in the fiducial model and simply varies the relative contributions from the core and later accreted pebbles, boulders and planetesimals. We consider scenarios where core mixing contributes 90, 70, 50 and 30\% of the total solids dissolved in Jupiter's envelope, and as Fig. \ref{fig:jupiter-grid}, the two intermediate cases provide good fits to observations, while the two extremes  deviate from observations. 

In summary, Jupiter's composition is only well reproduced if it obtained a large amount of solids from the outer solar system through core formation exterior to the N$_2$ and Ar snowlines, which in our disk model are placed at $\sim$26 and 40~au, respectively. Though note that the latter may becme revised inwards with new laboratory data made available (Schneiderman private comm.).  Once the core is formed, it is possible to reproduce all observed abundances if Jupiter either accreted its gaseous envelope in the outer solar system, or if it accreted its envelope at smaller radii together with a substantial amount of dissolvable solids in the form of pebbles and planetesimals. 

\section{Discussion} \label{sec:disc}

\subsection{Core formation and mixing}

We have showed that formation of Jupiter's core exterior to the N$_2$ and Ar snowlines, followed by core mixing into the gaseous envelope, provides a good fit to Jupiter volatile abundances when combined with inward migration, and with gas and further pebble and planetesimal accretion in the inner solar nebula.
This model hinges on several assumptions, 1) that it is possible to form a large, $\sim$20 Earth mass planetary core in the outer solar nebula, 2) that it can migrate to Jupiter's current location before the nebular gas dissipates, and 3) that a substantial portion of that core, $>$50\% could become dissolved in Jupiter's envelope. The plausibilities of these processes are is the subjects of this sub-section.

As introduced in \S\ref{sec:intro}, pebble accretion models that include a full dynamical treatment of the disk and nascent planets predict that Jupiter's core formed substantially further out in the solar system compared to Jupiter's present location \citep{Bitsch15,Ali-Dib17,Bitsch19}. How far out in the disk depends on when the core formed and on the disk mass and metallicity. The latter two are poorly constrained from solar system data and protoplanetary disk studies alike. Observations of protoplanetary disks may provide some information about when planet formation typically begins, however, assuming that observed sub-structures are associated with planet formation. The youngest disks that show substantial gaps are $<$1~Myrs old, including the iconic HL Tau disk \citep{Brogan15}. This suggests that the onset plane core formation is $<$1~Myr as well. \citep{Bitsch15} showed that the earlier Jupiter's core formed the further out in the disk it must have originated -- if it formed within a 1~Myr of the inception of the solar nebula and the nebula lasted for 3 Myrs, Jupiter's core likely formed beyond 30~au, in agreement with our model requirements. 

If Jupiter's core formed in the outer solar system, could it also have accreted its envelope at $>$30~au? Recent analyses of asteroid population data suggests that Jupiter's core was in place early in the history of the solar system \citep{Kruijer17}. This favors a scenario where the core alone formed at large distances and then migrated inwards. Recent work by \citet{Pirani19} also places Jupiter's core formation in the outer solar system, based on Trojan data, while most of the gas envelope is accreted closer to 5~au, consistent with our fiducial model.

The next issue is whether nitrogen and other elements accumulated during core formation would remain trapped in the core, or become well-mixed throughout Jupiter's interior after the planet's gas envelope was captured. Previous ab-initio calculations have shown that elements as heavy as Fe and Mg should be soluble in hydrogen at the pressures and temperatures expected in Jupiter's interior \citep{wilson2012rocky,wahl2013solubility, gonzalez2014ab}. Water ice is soluble at temperatures of around 3000~K at Jovian interior pressures \citep{wilson2012solubility}, which is much lower than the temperatures expected in the region of Jupiter's core. An initially icy core accreted in the outer solar system should therefore dissolve into the nearby hydrogen envelope. 

Although dissolution of heavy species from a core into a pure hydrogen envelope is expected, subsequent mixing into the gas envelope depends on convective processes that are still poorly understood under Jovian conditions \citep{leconte2013layered, nettelmann2015exploration,moll2017double}. Double-diffusive layered convection, if present, could reduce mixing efficiency, although its importance throughout Jupiter's evolutionary history is debated \citep{leconte2013layered,moll2017double,vazan2018jupiter}.  While the theory remains to be worked out there is recent empirical support for a core dissolution and dredging scenario: inter-comparison of interior models with Jupiter's low order gravitational moments J$_2$-J$_8$ measured by JUNO suggests a large, dilute core, which is consistent with a significant amount of core dredging having occurred \citep{wahl2017comparing}. To account for this \citet{Liu19} proposed that the young Jupiter collided head on with a large planet embryo, which shattered Jupiter's primordial core, and distributed its heavy elements into the inner envelope.

\subsection{The role of volatile entrapment}

So far all our models have assumed that volatile entrapment was unimportant in setting the bulk elemental abundances in the solar system. This is opposite to most previous explanations of Jupiter's enrichment pattern. We therefore briefly explore the outcome of our model when incorporating maximum entrapment assuming that $\sim$5 H$_2$O molecules are required for each entrapped hyper volatile, a 100\% entrapment efficiency, and that the hypervolatiles under consideration, CO, N$_2$ and noble gases, are all entrapped equally well. In other words we assume that an equal proportion of each hypervolatile is entrapped such that their sum does not exceed 20\% of the total number of water molecules. This likely over-predicts the amount of possible N$_2$ entrapment, since experiments show that N$_2$ is less efficiently trapped than CO  \citep[e.g.][]{Bar-Nun85,Bar-Nun07,Yokochi12}, and thus provides a limiting case for testing our conclusions. 
The assumption of that at least five water molecules are need for each entrapped hypervolatile is based on both clathration \citep{Lunine85} and amorphous ice entrapment studies \citep{Notesco03,Fayolle11,Simon19}. 

\begin{figure}
    \centering
    \includegraphics[scale=0.7]{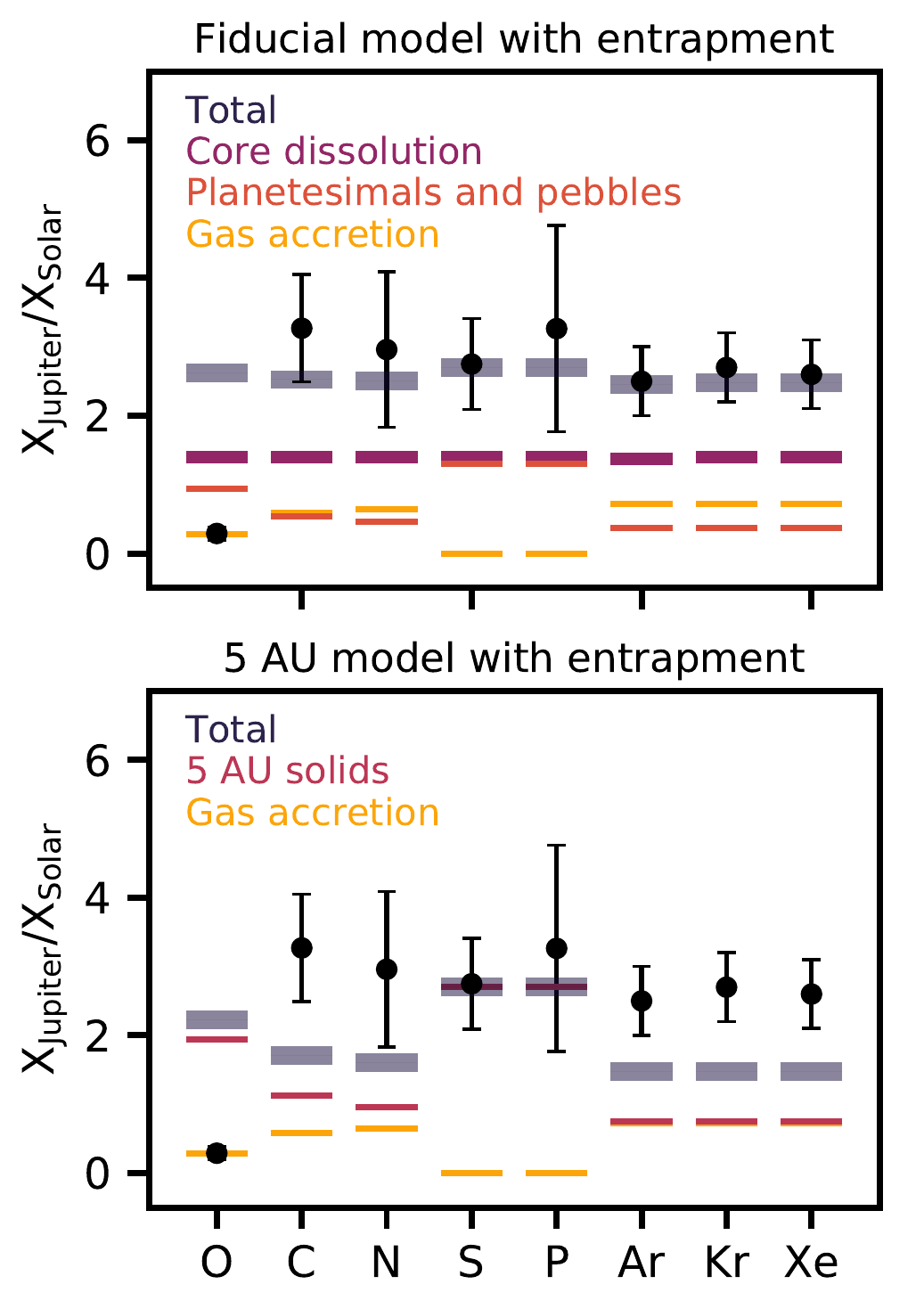}
    \caption{Jupiter envelope element enhancements over Solar in our fiducial model and in a model where Jupiter's core and envelope accretes at 5~au, assuming maximum entrapment of CO, N$_2$ and noble gases in water ice.}
    \label{fig:entrap}
\end{figure}

Based on the adopted water and hypervolatile abundances, a maximum of 28\% of the nebular hypervolatiles can become trapped. Figure \ref{fig:entrap} shows the results when incorporating this maximum level of entrapment for the fiducial model, and for a model where Jupiter completely forms at 5~au. The fiducial model results are basically unchanged by including entrapment. By contrast, in the model where Jupiter forms {\it in situ}, including entrapment does enhance N, C and noble gas abundances, but even our very optimistic, maximum entrapment assumption does not result in sufficient CO, N$_2$ or noble gases to explain observations. It is important to note that it is quite difficult to conceive of a scenario where maximum entrapment would occur, since N$_2$ is likely to freeze out on top of the H$_2$O ice matrix in the cooling Nebula rather than becoming perfectly mixed in with it, and should be trapped with a lower efficiency than the more abundant CO. 

\subsection{Predictions for Outer solar system missions}

The strongest prediction emerging from our model is  that the oxygen enhancement in Jupiter should be similar to that of C and N, i.e. that it is {\it not} substantially more enhanced. This is distinct from models that require entrapment of C and N carriers in H$_2$O ice, where oxygen will be enhanced by a substantially higher factor. 

A second set of predictions concerns Saturn. If Saturn and Jupiter formed at the same time, Saturn's inception was likely $\sim$3~au exterior to Jupiter's \citep{Pirani19}. In our model that places Saturn's formation well outside of the N$_2$ and Ar snowlines, and we should expect a similar enrichment pattern as in Saturn as in Jupiter, i.e. a near constant enrichment of O, C, N, S, P, Xe, Kr and Ar in Saturn's envelope. The formation timescale for both Saturn and Jupiter would be 2-3~Myrs and they would therefore have to start forming when the solar nebula was $<$1~Myr old \citep{Bitsch15}.
Another possible scenario is that Saturn formed after Jupiter had already fully assembled and migrated to its final position. In this scenario Saturn's core would begin to form when the solar nebula was already older than 2 Myrs, since it takes $\sim$2~Myrs for Jupiter-sized planet to form and migrate into place if its inception is beyond 30~au. Adopting a disk lifetime of 3~Myrs, Saturn would then need to form in $<$1~Myr, which limits its formation location to 15-20~au according to the model grid presented in \citet{Bitsch15}. In this scenario, P, O, C, Xe, and Kr should all be similarly enhanced in Saturn's envelope, while N and Ar should be underabundant. Better constraints on Saturn's elemental composition is thus key to constrain when and how the outer Solar System assembled.

\section{Conclusions} \label{sec:conc}

Jupiter's near-uniform enhancements in C, N, S, P, Ar, Kr and Xe are difficult to explain if Jupiter formed close to its current location at 5~au from the Sun. At these radii solids are expected to be depleted in nitrogen, carbon, and noble gases compared to oxygen, sulphur and phosphorous, and Jupiter's composition cannot then be explained by accretion of locally assembled solids. Transport of solids from the outer solar system might provide a partial answer, but to fit observations N$_2$ entrapment in water ice would have to be near-complete in the outer solar nebula, and locally assembled pebbles and planetesimals must have been prevented from polluting Jupiter's envelope with O-rich solids. Both seem unlikely, but a final test will come with Juno's measurement of oxygen in Jupiter's envelope. If the observed nitrogen enhancement is due to N$_2$ entrapment in water, the oxygen enrichment in Jupiter should be high, since at least five and more likely ten water molecules are required for each entrapped N$_2$ molecule.

We propose that Jupiter's abundances are instead due to that Jupiter's core formed in the cold ($<$25~K), outer solar nebula, beyond the N$_2$ and Ar snowlines. At these radii ($>$30~au) solids contained Solar ratios between O, C, N, S, P and noble gases. During envelope accretion and later planetesimal bombardment a substantial fraction of the primordial core was dissolved into Jupiter's envelope, producing the characteristic abundance pattern. Based on a small set of toy models, this scenario is robust to later solid accretion close to Jupiter's current feeding zone during e.g. the clean-up phase of planet formation, as long as a majority of solids dissolved in Jupiter's envelope originated beyond the N$_2$ and Ar snowlines. A key prediction of this model is that oxygen should be enhanced at a similar level to carbon and nitrogen in Jupiter's envelope.

We note that our proposed formation location for Jupiter's core is consistent with recent  pebble accretion models, which also place Jupiter's inception in the outer solar nebula, and with observations of an extended core in Jupiter. It also fits with increasing evidence of planet-induced sub-structure at 10s of au in many protoplanetary disks, suggesting that gas giants may commonly begin their existence at 10s of au, followed by inward migration during their early stages of formation.  

\acknowledgments

 K.I.\"O. acknowledges support from the Packard Foundation, and R. W. acknowledges support from NASA Habitable Worlds grant NNX16AR86G.

\bibliographystyle{aasjournal}

\end{document}